\documentclass[aps,prb,twocolumn,floats,showpacs,superscriptaddress]{revtex4}
\usepackage{graphicx,epsfig,psfrag}
\usepackage{times}
\usepackage{graphics,dcolumn,bm,epic,eepic,float}
\usepackage{amssymb,amsmath,rotate,color}
\usepackage{graphicx}

\begin{document}
\unitlength 1 cm
\newcommand{\be}{\begin{equation}}
\newcommand{\ee}{\end{equation}}
\newcommand{\bearr}{\begin{eqnarray}}
\newcommand{\eearr}{\end{eqnarray}}
\newcommand{\nn}{\nonumber}
\newcommand{\la}{\langle}
\newcommand{\ra}{\rangle}
\newcommand{\cd}{c^\dagger}
\newcommand{\vd}{v^\dagger}
\newcommand{\ad}{a^\dagger}
\newcommand{\bd}{b^\dagger}
\newcommand{\tk}{{\tilde{k}}}
\newcommand{\tp}{{\tilde{p}}}
\newcommand{\tq}{{\tilde{q}}}
\newcommand{\eps}{\varepsilon}
\newcommand{\vk}{{\vec k}}
\newcommand{\vp}{{\vec p}}
\newcommand{\vq}{{\vec q}}
\newcommand{\vkp}{\vec {k'}}
\newcommand{\vpp}{\vec {p'}}
\newcommand{\vqp}{\vec {q'}}
\newcommand{\bk}{{\bf k}}
\newcommand{\bp}{{\bf p}}
\newcommand{\bq}{{\bf q}}
\newcommand{\br}{{\bf r}}
\newcommand{\bR}{{\bf R}}
\newcommand{\up}{\uparrow}
\newcommand{\down}{\downarrow}
\newcommand{\fns}{\footnotesize}
\newcommand{\ns}{\normalsize}
\newcommand{\cdag}{c^{\dagger}}
\newcommand{\lc}{\langle\!\langle}
\newcommand{\rc}{\rangle\!\rangle}

\title{Stable local moments of vacancies and hollow-site impurities in graphene}
\author{M. Mashkoori}
\affiliation{Department of Physics, Sharif University of Technology, Tehran 11155-9161, Iran}
\author{S. A. Jafari}
\affiliation{Department of Physics, Sharif University of Technology, Tehran 11155-9161, Iran}
\affiliation{Center of excellence for Complex Systems and Condensed Matter (CSCM), Sharif University of Technology, Tehran 1458889694, Iran}
\affiliation{School of Physics, Institute for Research in Fundamental Sciences (IPM), Tehran 19395-5531, Iran}

\begin{abstract}
Taking into account the possibility of a p-wave hybridization function $V(\vk)$ of ad-atom with Dirac electrons in
graphene -- which arises for vacancies and hollow-site impurities --
we study the nature of magnetic moment formation within the single impurity Anderson model (SIAM).
Compared to the s-wave hybridization function, we find that 
the local moments formed within the Hartree mean field are robust against the change in
the parameters of the model. Further we investigate the stability of the local moments
with respect to quantum fluctuations by going beyond the Hartree approximation.
We find that for parameter regimes where local moments formed by top-site ad-atoms
are completely washed out by quantum fluctuations, those formed by vacancies (or hollow-site 
impurities) survive the quantum fluctuations captured by 
post-Hartree approximation. Hence vacancies
and hollow-site ad-atoms  are suitable candidates to produce stable local moments.
\end{abstract}

\maketitle

\section{Introduction}
Since its isolation in the laboratory, graphene has stimulated extensive research
activity among condensed matter and material physicists~\cite{Novoselov1,Novoselov2,Kim}.
Distinct features of graphene are its two dimensional structure and the 
nature of low-energy excitations which are described by the Dirac equation~\cite{NetoRMP}.
These two characteristics lead to fascinating properties of graphene.
For example two dimensionality allows for tuning the carrier type density
via a gate voltage as well as the possibility of adding/substituting
various atoms on graphene~\cite{Danny}. The two dimensional structure further
allows to conveniently remove some of the carbon atom to
create vacancies which are responsible for spin-half magnetic states~\cite{Nair}.
The potential distortion resulting from the creation of a vacancy
can in turn lead to a shallow impurity state which can hybridize
with $2p_z$ orbitals of three neighboring carbon atoms. Assumption of such
a hybridization, after Fourier transformation from localized Wannier states
to corresponding Bloch wave-functions results in a momentum dependence
in the hybridization matrix element $V(\vk)$ between the state localized on vacancy 
and the $\pi$-bands of graphene, which after linearization
around the Dirac points acquires at low-energies the functional form 
of a p-wave dependence on momentum~\cite{Uchoa2,JafariTohyama1}:
\be
  V(\vk)=(k_x-ik_y) \tilde V
  \label{vk.eqn}
\ee
The above form is consistent with the fact that the pseudo-spin structure
of the Bloch wave functions in graphene allows them to hybridize with 
external states in both $l=0$ and $l=1$ angular momentum channels~\cite{Sengupta}. 
The above form also holds for the hybridization between the localized states
of an ad-atom in a hollow-site position which hybridizes with {\em both} sublattices
via $V(\vk)$ and $V^*(\vk)$ functional forms. In this respect hollow-site 
ad-atoms and vacancies differ from top-site impurities in that they have
p-wave hybridization, while the later has s-wave hybridization.

The phase diagram of magnetic states in graphene has been previously studied by
Uchoa and coworkers~\cite{Uchoa}. They studied the $l=0$ channel~\cite{Sengupta}
corresponding to momentum-{\em independent} hybridization function. 
Within the Hartree mean field theory they obtained a phase
diagram for the magnetic states which significantly differs from the corresponding
one for ordinary metals~\cite{Anderson}. In this work we focus on the momentum 
dependent hybridization function of the p-wave form relevant to 
vacancies and hollow-site ad-atoms, and investigate the local spectral properties of
the SIAM both within and beyond the Hartree mean field. Within the mean field, 
in contrast to the momentum-independent case~\cite{Uchoa}, we find that magnetic states 
are formed in much larger region of the parameter space. This could be 
interpreted as the parametric robustness of the local moments arising from
vacancies or hollow-site ad-atoms. Then within the equation of motion approach
we proceed one step beyond the Hartree decoupling and find that the
local moments due to p-wave hybridization display stability against
quantum fluctuations beyond the mean field theory.
These findings shed light on the experimentally observed formation of 
spin-half states which are attributed to vacancies in graphene~\cite{Nair}.

\begin{figure}[b]
\center
\includegraphics[width = 2.5cm]{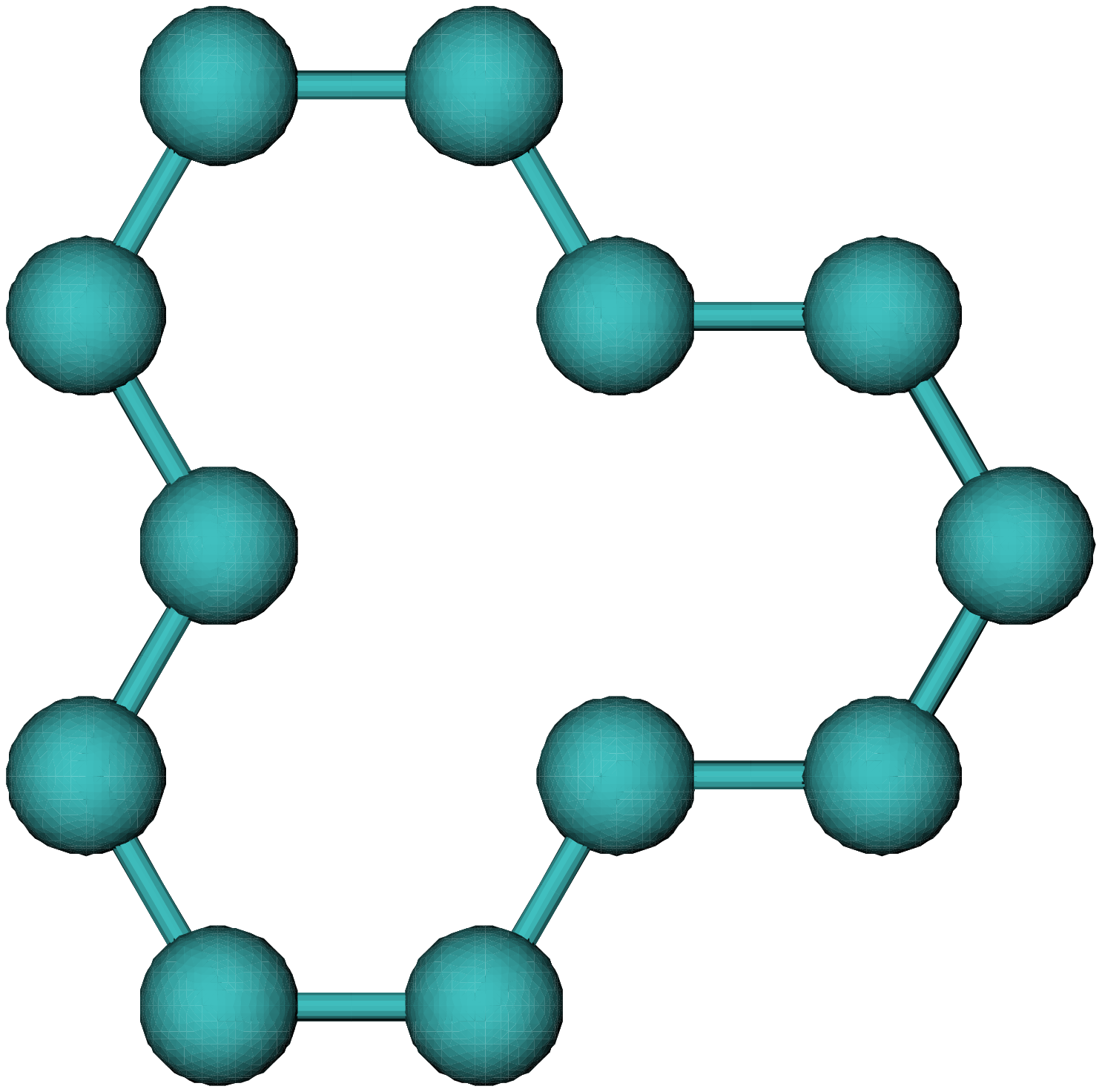}
\includegraphics[width = 2.5cm]{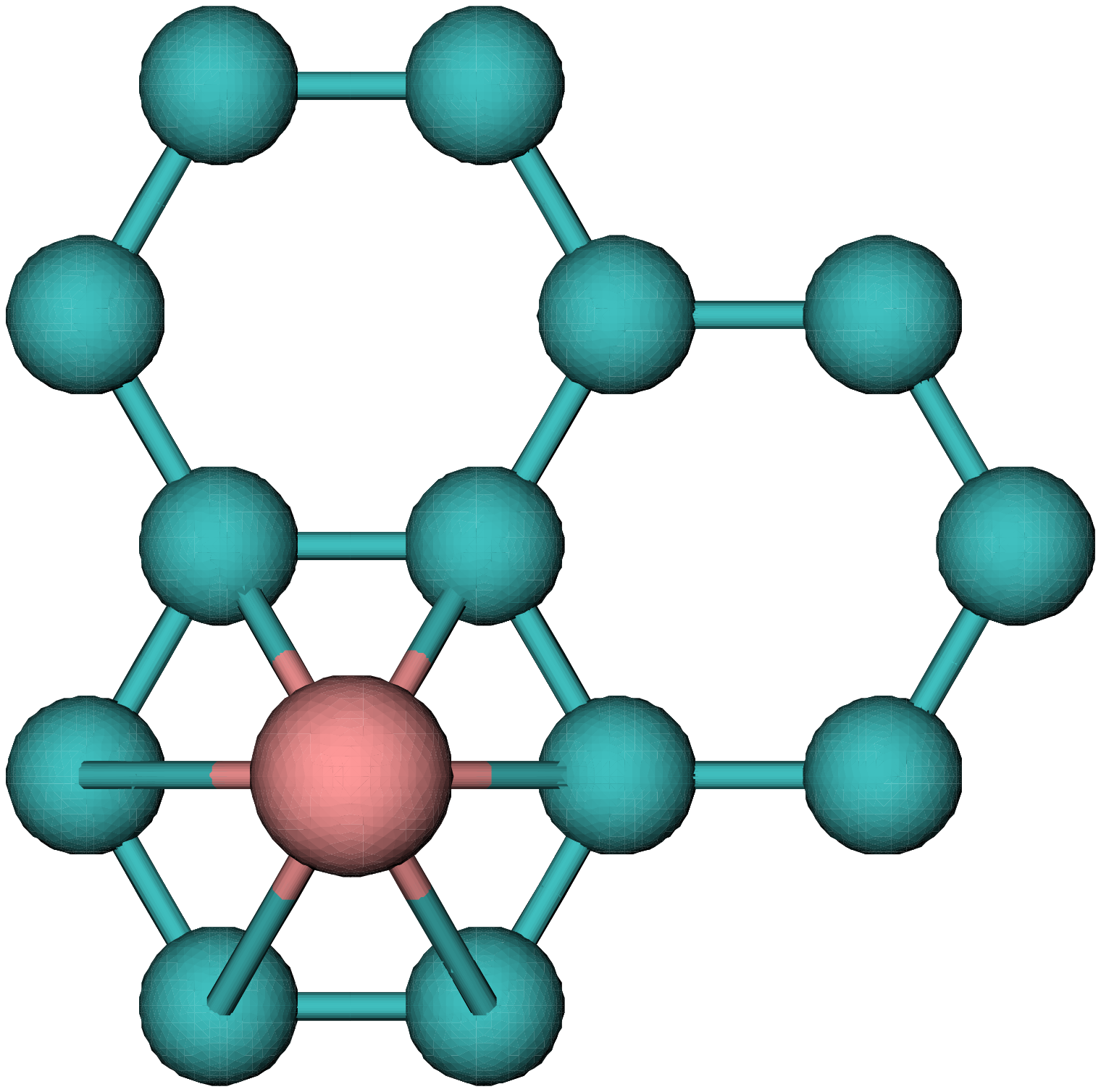}
\includegraphics[width = 2.5cm]{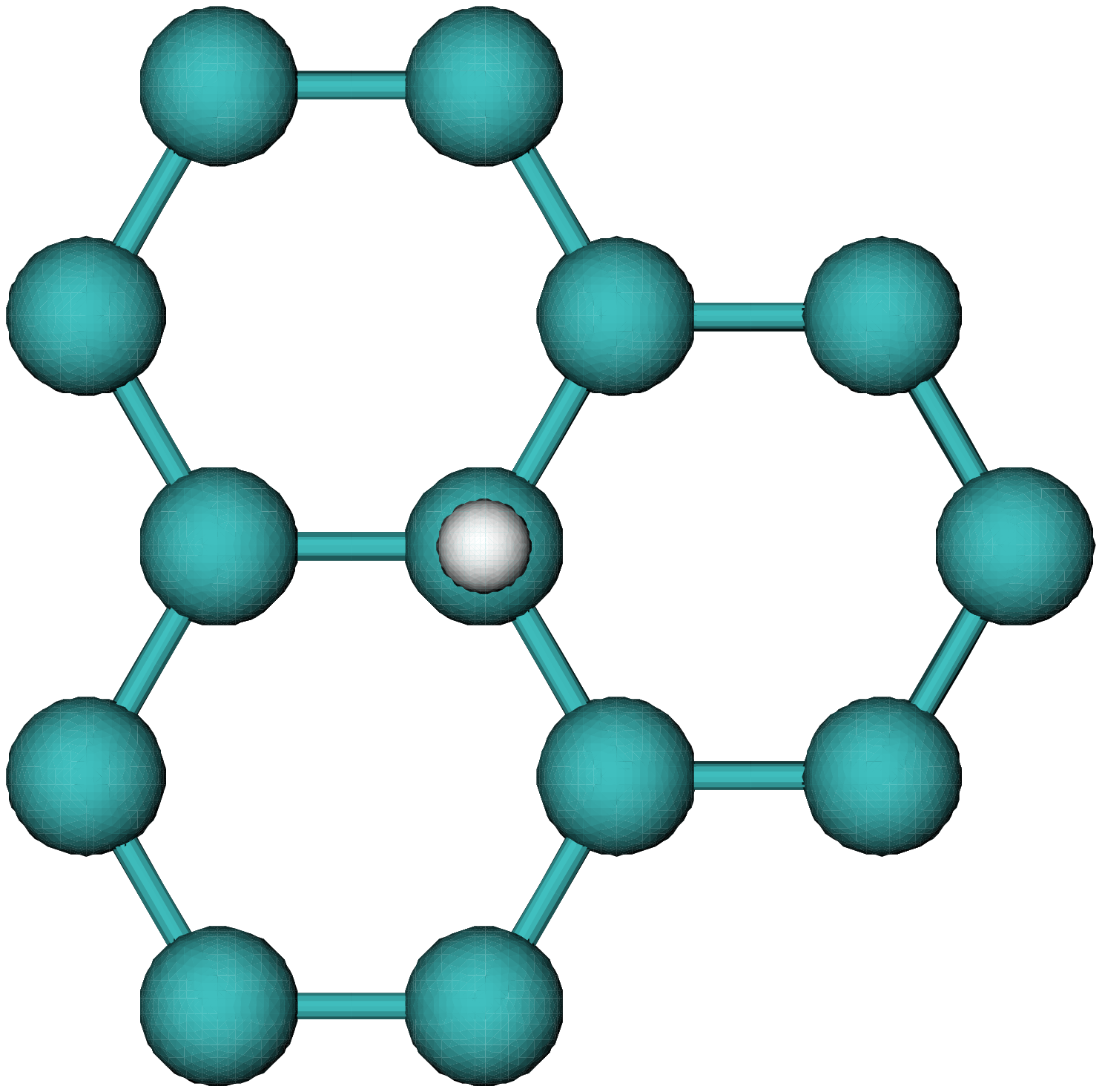}
\vspace{-8mm}
\caption{Vacancy (left), hollow site ad-atom (middle) and top-site 
ad-atoms (right) in graphene. The impurity state centered around the vacancy 
hybridizes with three carbon atoms form a sublattice, while the hollow site
ad-atom in addition hybridizes with another three atoms from the
other sublattice. The impurity orbital associated with the top-site 
impurity hybridizes only with the $2p_z$ orbital of one underneath carbon atom.
}
\label{vh.fig}
\end{figure}

\section{Formulation of the problem}
The Anderson model for graphene with a vacancy can be written as,
	\[H = {H_{\rm g}} + {H_{\rm hyb}} + {H_{\rm imp}}\] 	
The first term of above Hamiltonian describes ${\pi}$ electrons of graphene. 
The second term describes hybridization of impurity state  with Dirac fermions in 
graphene. To create a vacancy  we assume that a carbon atom from A-sublattice 
has been removed. These terms after a standard Hartree mean field factorization will 
be written as
\begin{equation}
\begin{array}{l}
   {H_{\rm g}} =  - t\sum\limits_{\vk\sigma } {(\phi( \vk)a_{\vk\sigma }^\dag {b_{\vk\sigma }} + {\phi ^*}( \vk )b_{\vk\sigma }^\dag {a_{\vk\sigma }})} \\
{H_{\rm hyb}} = \frac{1}{\sqrt N}\sum\limits_{\vk\sigma } V(\vk) b_{\vk\sigma }^\dag {d_\sigma } + h.c.\\
{H_{\rm imp}} = \sum\limits_\sigma  {{\varepsilon _\sigma }d_\sigma^\dag {d_\sigma }} 
~~~~;{\varepsilon _\sigma } \equiv {\varepsilon_0} + U\langle {{n_{\bar \sigma }}} \rangle 
\end{array}
\label{ham.eqn}
\end{equation}
where $\phi(\vk)=\sum_{n=1}^3\exp(-i\vec \delta_n.\vk)$ is the form factor associated with
three neighbors connected by vectors $\delta_n$, with $n=1,2,3$. The creation operators
$a^\dagger_{\vk\sigma}$ and $b^\dagger_{\vk\sigma}$ create Bloch electrons on A and B sublattices 
with spin $\sigma$, respectively and $d^\dagger_\sigma$ creates an electron with spin $\sigma$ in 
the impurity state associated with the vacancy. Here $\eps_0$ is the impurity level, which after
incorporating the Hubbard repulsion $U$ splits into two sub-levels denoted by $\eps_\sigma$.
The hybridization function $V(\vk)$ as emphasized in Eq.~\eqref{vk.eqn} is proportional to 
the form factor $\phi(\vk)$ which in the continuum limit becomes proportional $k_x-ik_y$.

After the Hartree mean field factorization implicit in Eq.~\eqref{ham.eqn},
the impurity Green function can be written as,
\be
  G_{d\sigma}^{-1} (\omega)= {\omega  - {\varepsilon _\sigma } - \Sigma^{\Delta}_d(\omega)} 
  \label{dyson.eqn}
\ee
where,
\be
   \Sigma^{\Delta}_d(\omega) = \sum_\vk |V(\vk)|^2 \frac{\omega}{\omega^2-t^2|\phi(\vk)|^2}
\ee
The above integral can be easily evaluated to give,
\be
 - \frac{{\omega\Delta  }}{{\pi {t^2}}}\left[ {{D^2} + {\omega ^2}\ln ( {\frac{{\vert {{\omega ^2} - {D^2}} \vert}}{{{\omega ^2}}}} ) + i\pi \omega \vert \omega  \vert\theta ( {D - \vert \omega  \vert} )} \right]\\
\label{selfen} 
\ee
where $\Delta  \equiv \frac{{\pi {V^2}}}{{{D^2}}}$, where $D\sim 7$eV is the high-energy
band cutoff chosen such that the total number of electrons in the linearized
band are the same as $\pi$-band.
Note that in contrast to the momentum-independent case -- see. Eq.~\eqref{sfs.eqn} --
where the imaginary part
of the $\Sigma^\Delta_d$ is proportional $|\omega|$, here the p-wave momentum-dependence
of the hybridization gives rise to anomalous $|\omega|^3$ dependence in the
imaginary part of $\Sigma^\Delta_d$. 
Inserting the above equation in~\eqref{dyson.eqn}
the impurity Green function becomes,
\begin{equation}
   G_{d\sigma}(\omega) = \frac{1}{{{Z^{-1}}( \omega)\omega - {\varepsilon_\sigma } 
   +i|\omega|^3\Delta \theta(D-|\omega|)/t^2 }},
\end{equation}
where $ Z^{-1} ( \omega ) $ is given by,
\begin{equation}
{Z^{ - 1}}( \omega  ) = 1 + \frac{{{V^2}}}{{{t^2}{D^2}}}\left[ {{D^2} + {\omega ^2}\ln ( {{{\vert {{\omega ^2} - {D^2}} \vert} \mathord{\left/
 {\vphantom {{\vert {{\omega ^2} - {D^2}} \vert} {{\omega ^2}}}} \right.
 \kern-\nulldelimiterspace} {{\omega ^2}}}} )} \right].
\end{equation}
The imaginary part of ${\Sigma^{\Delta}_d}(\omega)$ shows the broadening of localized level due to hybridization.
Here, in contrast to normal metals, the broadening displays strong $\omega$-dependence. 
The $\omega$-dependence in the p-wave case is even stronger than the s-wave case.
Then the local spectral function becomes,
\begin{equation}
  A_{d\sigma} (\omega) = \frac{\Delta }{{\pi {t^2}}}\frac{{{{\vert \omega  \vert}^3}\theta ( {D - \vert \omega  \vert} )}}{{{{( {{Z^{ - 1}}( \omega  )\omega  - {\varepsilon _\sigma }} )}^2} + {{{\Delta ^2}{\omega ^6}} \mathord{\left/
 {\vphantom {{{\Delta ^2}{\omega ^6}} {{t^4}}}} \right.
 \kern-\nulldelimiterspace} {{t^4}}}}}
 \label{eq9.eqn}
\end{equation}
which in turn gives the occupation of $\sigma$ sub-band as,
\begin{equation}
	{n_\sigma } = \frac{\Delta }{{\pi {t^2}}}\int_{ - \infty }^\mu  {d\omega \frac{{{{\vert \omega  \vert}^3}\theta ( {D - \vert \omega  \vert} )}}{{{{( {{Z^{ - 1}}( \omega  )\omega  - {\varepsilon _\sigma }} )}^2} + {{{\Delta ^2}{\omega ^6}} \mathord{\left/
 {\vphantom {{{\Delta ^2}{\omega ^6}} {{t^4}}}} \right.
 \kern-\nulldelimiterspace} {{t^4}}}}}}
 \label{sc.eqn}
\end{equation}
In this equation, the $n_\up$ is given as an integral involving
$\eps_\up$ which itself depends on $n_{\down}$ as given by Eq.~\eqref{ham.eqn}.
These equations for $n_\up$ and $n_\down$ must be solved self-consistently. 
Eq.~\eqref{eq9.eqn} must be compared to Eq. (9) of Ref.~\cite{Uchoa} 
corresponding to s-wave hybridization of top-site impurities.
In order to make the comparison, we have introduced function $Z^{-1}(\omega)$ along with the 
notations of this reference. The function $Z^{-1}(\omega)$
in our case differs from the $\vk$-independent hybridization in two respects: (i) the 
logarithmic part of the $\omega$-dependence has acquired an anomalous $\omega^2$
factor compared to $\vk$-independent case. (ii) Comparing $\Sigma^\Delta_d$ of the
s-wave case -- Eq.~\eqref{sfs.eqn} -- in the p-wave case an additive $D^2$ term appears.
To complete the self-consistency cycle, one needs to accurately evaluate the integrals
connecting $n_\up$ and $n_\down$. The authors of Ref.~\cite{Uchoa} argue that, neglecting
the $\omega$-dependence of $Z^{-1}(\omega)$ allows for analytic evaluation of the 
integral at the cost of introducing a few percent error. But as we 
will show in the sequel, since such errors are repeated several times through out the
self-consistency cycle, the final result are sensitive to the $\omega$-dependence
of the $Z^{-1}(\omega)$ function.

\section{Numerical results}
In this section we present numerical results for the self-consistent solution at 
the level of Hartree mean field equations. To begin with, 
we revisit the top-site hybridization
case discussed in Ref.~\cite{Uchoa} and show how the approximation of neglecting
the $\omega$-dependence of $Z^{-1}(\omega)$ leads to qualitatively different phase diagram.
Then we focus on the case of p-wave hybridization function pertinent to vacancy and
hollow-site impurities. 

\subsection{Local moments from top-site impurities}
In this case the hybridization $V_\vk$ does not depend on $\vk$, and the 
occupation of spin $\sigma$ impurity sub-band is given by~\cite{Uchoa},
\begin{equation}
   n_{\sigma}=-\frac{1}{\pi}\int^{\mu}_{-D}d\omega 
   \frac{\Delta\vert\omega\vert\theta(D-\vert\omega\vert)}{[Z^{-1}(\omega)\omega-\varepsilon_{\sigma}]^2+\Delta^2\omega^2},
\end{equation}
where
\begin{equation}
   Z^{-1}(\omega)= 1 + \frac{V^2}{D^2} \ln(\frac{\vert D^2 - \omega^2 \vert}{\omega^2}) .
\end{equation} 
Here {\em we do not apply the approximation} 
$Z^{-1}(\omega)\approx Z^{-1}(\varepsilon_{\sigma})$ 
{\em and retain the full energy dependence of} $Z^{-1}$. The price one has to pay is to do
the integrations necessary to get $n_\sigma$ numerically. 
We define two standard variables $X=D\Delta/U$ and $Y=(\mu-\eps_0)/U$ in terms of
which the phase diagram is traditionally constructed~\cite{Anderson}.
\begin{figure}[t]
\center
\includegraphics[width = 8.0cm]{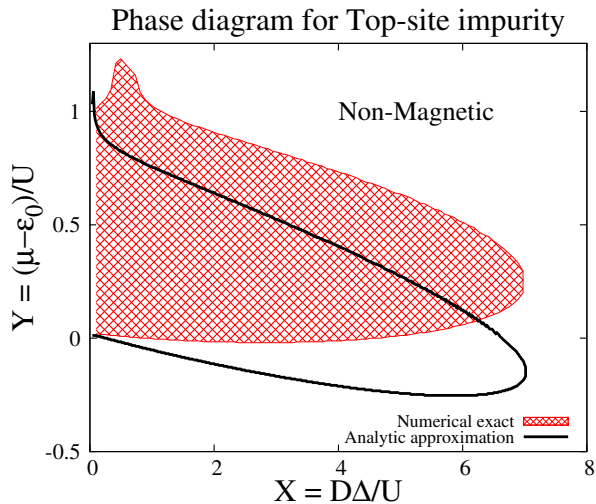}
\caption{(Color online) Phase diagram for the local moment formation in SIAM for top-site impurity 
in graphene.The area enclosed by the curves and the $Y$-axis is magnetic. In this plot the variables 
$X$ and $Y$ are defined by $X = \frac{\Delta D}{U}$ , $Y=\frac{(\mu-\varepsilon_{0})}{U}$. 
Other parameters are $\frac{V}{D} = 0.14 $ and $\frac{\varepsilon_{0}}{D}=0.043$.
The black thick solid line corresponds to the approximate analytic integration 
of Ref.~\cite{Uchoa}. The red (filled) plot corresponds to numerically exact evaluation
of integrals.
}
\label{pd.fig}
\end{figure}
The result is show in Fig.~\ref{pd.fig}. The black (thick solid) plot 
corresponds to the approximate solution where the energy dependence of $Z^{-1}$ 
has been neglected.
In the red plot (filled plot) retaining the full energy dependence of $Z^{-1}$ 
the integrals are evaluated
numerically. As can be seen, there are two major differences: (i) The numeric result compared to the 
approximate one shows a significant shift of the lower lobe of the phase 
boundary to positive $Y$ values 
corresponding to $\mu>\varepsilon_0$. In this respect for the top-site impurity states
in graphene magnetic states are dominantly in $\mu>\eps_0$ region, 
akin to normal metals. The asymmetry of the graph around $Y=0.5$ however 
is the feature distinct from normal metals and upon precise calculation
of integrals this asymmetry will be even more enhanced for the same
set of parameters indicated in the figure caption.
(ii) The second qualitative difference is that the re-entrant 
behavior seen as a hump for small values of $X$ will be more pronounced when the integrals
are evaluated exactly. Although neglecting the energy dependence in $Z^{-1}$ leads
only to small error in {\em each} evaluation of the integral, 
repeating this procedure in the self-consistency process, propagates and enhances 
the errors. With this point in mind,
in the following we avoid using such approximations and focus on the self-consistent
solution of the mean field equations for the p-wave hybridization function.
\begin{figure}[t]
   \center
   \includegraphics[width= 8.0cm]{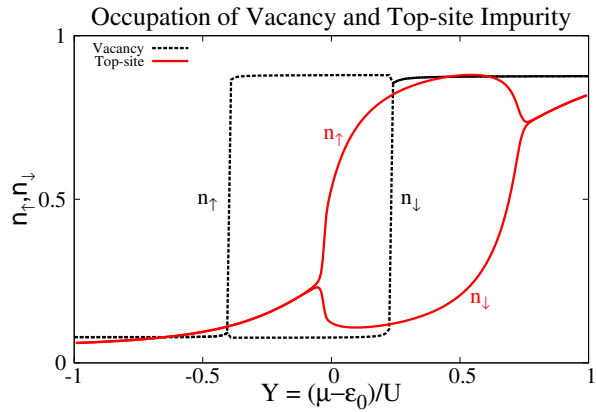}
   \caption{(Color online) Comparison of the self-consistent mean field occupations $n_\up$ 
   and $n_\down$ as a function of $Y$ for a fixed value $X=5.0$ for the vacancy (black dashed line)
   and top-site impurity (red solid line). The region of local moment formation for vacant 
   graphene is shifted towards negative values of $Y$ in comparison to top-site case.
   Moreover within the whole region, magnetic moment formed via p-wave hybridization 
   is more or less constant and suddenly drops to zero at the phase boundary, while in the case of
   top-site impurity, the moment smoothly vanishes by approaching the boundary.
   }
   \label{top-hollow.fig}
\end{figure}

\subsection{Local moments from vacancies or hollow-site impurities}
Now we proceed by self-consistently solving the mean field equations~\eqref{eq9.eqn}
for vacancy induced local moments. In Fig.~\ref{top-hollow.fig} we 
compare the self-consistently determined values of occupation numbers
$n_\up$ and $n_\down$ for a fixed value of $X=5.0$
in the two cases corresponding to top-site impurity and vacancy.
For very small 
and very large values of $Y$ the $n_\up$ and $n_\down$ curves coincide
which means the magnetic moment is zero. For intermediate values of 
$Y$ the curves corresponding to $n_\up$ and $n_\down$ split. The 
amount of splitting determines the magnetic moment. 
 As can be seen the local moment formation region in vacancy sets in
smaller values of $Y$ on the negative values, while in the top-site
case, the local moment region sets in $\mu-\varepsilon_0\lesssim 0$.
Moreover in the case of vacancy the magnetic moment inside the 
local moment region stays more or less constant and abruptly 
vanishes by approaching the boundary, while in the top-site case,
the local moment smoothly vanishes as the boundary is reached. 

\begin{figure}[t]
\center
\includegraphics[width= 8.0cm]{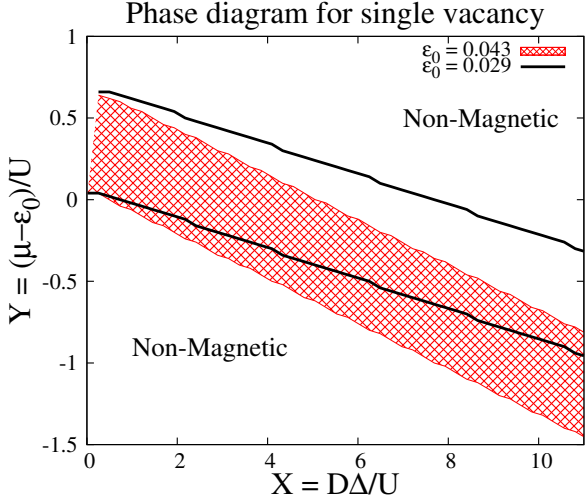}
\caption{ (Color online) Local moment phase diagram of the single vacancy Anderson model.
The variables $X$ and $Y$ are defined by 
$X = \frac{\Delta D}{U}$ , $Y=\frac{(\mu-\varepsilon_{0})}{U}$. 
The parameters in both plots are  $\frac{V}{D} = 0.14 $ while their impurity energy levels 
are different, with filled (red) plot corresponding to $\frac{\varepsilon_{0}}{D}=0.043$ and
thick solid (black) line corresponding to $\frac{\varepsilon_0}{D}=0.029$.
}
\label{pdv.fig}
\end{figure}

Repeating the above analysis for other values of $X$ we obtain the phase diagram depicted
in Fig.~\ref{pdv.fig}. This plot has been generated for $V/D=0.14$ and two 
different values of $\eps_0/D=0.029$ and $\eps_0/D=0.043$ as indicated in the figure. 
The plots corresponding to $\eps_0/D=0.043$ in this figure can be compared to the one
in Fig.~\ref{pd.fig} of the top-site impurities as they correspond to the same set of parameters.
As can be noted by comparing Fig.~\ref{pd.fig} and Fig.~\ref{pdv.fig}, in the case of
p-wave hybridization, the local magnetic moments are found in a much
larger region of parameter space. This can be interpreted as the parametric robustness
of local moments arising from vacancies or hollow-site impurity states
compared to those due to top-site impurity states. 
Another important aspect can be noticed by comparison of the p-wave phase
diagram to the numerically exact plot in Fig.~\ref{pd.fig} of s-wave case
is that in the later case the local moment region dominantly lies in the positive values of $Y$,
a feature also shared by normal metals, while in the former case, the local moment
region is substantially extended to negative values of $Y$ and has been elongated
as an stripe shaped region. Such elongation has been reported 
in a similar study for bilayer graphene~\cite{Heidarian} and has been shown to
arise from a large contribution in the real part of the self-energy in the vacant case 
which arises from a constant shift on the scale of bandwidth $D$. Such self-energy shifts
can be absorbed into $\varepsilon_0$ defining a largely shifted effective $\tilde \varepsilon_0$.
This is consistent with the fact that small changes in $\varepsilon_0$ does not change the
elongated nature of the phase diagram. Moreover up to quite large values of
$X\sim 12$ -- which is already much larger than physically conceivable values 
for physical graphene -- the width of the range of $Y$ values for which the local 
moment is formed does not vary much. 
The larger magnetic moment formation area in the phase space can be interpreted as
the robustness of the local moments formed by vacancies compared to top-site impurities,
and is consistent with the observed spin-half magnetic moment in 
vacant graphene~\cite{Nair}.

\subsection{Internal structure of the impurity orbital in hollow-site ad-atoms}
Large ad-atoms prefer the hollow-site position when added to a graphene sheet~\cite{Cohen}
while small atoms such as hydrogen prefer the top-site position~\cite{Danny}. 
Therefore implicit assumption of s-wave {\em orbital structure for the impurity atom itself} 
is only feasible for top-site positions. With larger ad-atoms in the 
hollow-site positions, higher multiplicities and more complex 
symmetry patterns may appear. 
In this section we address the question local moment formation for impurity 
orbitals that preserve the time reversal symmetry within the 
Hartree mean field theory.
For the case of hollow-site ad-atom, the s-wave symmetry is formally identical
to what we considered in the vacancy case and that is why we treated these
two cases on the same footing. The only difference between them is that the $\Sigma^\Delta_d$ 
in the s-wave hollow-site case is twice the vacancy case. The reason is simply
related to the fact that now in addition to three nearest neighbors from a given
sublattice, three nearest neighbors from the other sublattice also emerge. 
Hence the phase diagram of local moments in the case of s-wave ad-atom in the
hollow-site position is identical to that of vacancy after a re-scaling of
hybridization strength $V$, i.e.: $V^{\rm hollow-site}=\sqrt 2 V^{\rm vacancy}$.

The $p$ orbitals alone can not form a linear combination compatible with the
six-fold rotational symmetry of graphene. The linear combination of $d$ orbitals
compatible with the symmetries of honeycomb hexagon requires complex coefficients
which breaks the time reversal symmetry. Therefore the simplest non-trivial
higher angular momentum pattern compatible with the rotational symmetry of the
honeycomb lattice is the $f$-wave pattern. 
For this purpose we consider in the hollow-site configuration a hybridization 
Hamiltonian of the following form,
\begin{align} 
   {H_{\rm hyb}} = \frac{V}{\sqrt N}\sum_{\vk\sigma} {
   ( \phi(\vk) b_{\vk\sigma }^\dag + e^{i\alpha}{\phi ^*}(\vk) a_{\vk\sigma }^\dag  ) d_\sigma + h.c.}
\end{align}
In above Hamiltonian, when the parameter $\alpha$ takes the value $\pi$ it corresponds to 
f-wave pattern of the atomic orbital. Obviously $\alpha=0$ reduces to the s-wave atomic orbital
where the impurity is hybridized with both A and B sublattice equivalently, while in f-wave case 
the hybridization pattern changes the sign of $V$ after each $2\pi/6$ rotation.

The equation of motion for the Green function of the above Hamiltonian will now depend on the 
phase $\alpha$,
\begin{equation}
G_{d\sigma}^{-1}(\omega) = \omega	- \varepsilon_{0} - \Sigma^{\Delta}_{d}(\omega)
\end{equation}
where the dependence on $\alpha$ is through the self energy:
\begin{equation}
\begin{array}{c}
\Sigma^{\Delta}_d(\omega) = {V^2}\sum\limits_\vk \frac{\omega}{\omega^2-t^2|\phi(\vk)|^2}(g(\vk,\omega)+e^{-2i\alpha}g^*(\vk,\omega)),\\
g^*(\vk,\omega) \equiv \frac{{v_F^2{k^2}}}{{{t^2}}}( {1 - {{{v_F}k{e^{i(3\theta  + \alpha )}}} \mathord{\left/
 {\vphantom {{{v_F}p{e^{i(3\theta  + \alpha )}}} \omega }} \right.
 \kern-\nulldelimiterspace} \omega }} ).
\end{array}
\label{hself.eqn}
\end{equation}
For both $\alpha$ values of zero and $\pi $, we have ${e^{ - 2i\alpha }}=1$. 
Furthermore the $\theta$ integration of $e^{i(3\theta+\alpha)}$ produces zero.
Therefore the self energy at the Hartree level will not depend on $\alpha$,
and the difference between s-wave and f-wave internal structures of the
impurity orbital does not appear at the mean field level. 


\section{Stability of mean-field approximation}
So far we have investigated within the mean field level the formation
of local moments for two situations corresponding to the s-wave and
p-wave hybridization function $V(\vk)$. We have further checked that
in the hollow site case the internal structure of the impurity orbital
itself does not matter within the mean field theory. 
When the local moments are formed in the Hartree mean field level,
the question will be, what happens to the local magnetic moments
when quantum fluctuations beyond the mean field are taken into account.
A proper treatment of the effect of fluctuations leads to the
dynamical screening of the magnetic moment (Kondo screening). In this section we
address this question by going one step beyond the Hartree mean field
within the equation of motion approach. To do this we use the 
standard notation for the Fermionic correlation functions:
\begin{equation}
\langle{\langle{f_\sigma(t)|f_\sigma^\dagger(t^\prime)}\rangle}\rangle = -i\theta(t-t^\prime)
\langle{\lbrace f_\sigma(t)|f_\sigma^\dagger(t^\prime)\rbrace}\rangle.
\end{equation}
Writing the equation of motion for the above Green function in frequency domain gives,
\begin{equation}
 (\omega  - \varepsilon _0 - \Sigma^{\Delta}_d(\omega))\langle{\langle{{f_\sigma }|f_\sigma ^\dag}\rangle}\rangle  = 1 + U\langle{\langle{{f_\sigma}{n_{\bar\sigma}}|f_\sigma ^\dag}\rangle}\rangle.
\label{eqom}
\end{equation}
In the Hartree approximation the correlation function on the right of Eq.~\eqref{eqom} 
is approximated by $ \langle{n_{\bar\sigma}}\rangle\langle{\langle{f_\sigma}|f_\sigma ^\dag}\rangle\rangle$,
which closes the equations, and the effect of Hubbard $U$ will be to replace 
$\eps_0\to\eps_\sigma=\eps_0+U \langle n_{\bar\sigma}\rangle$. But here we do not close 
the equations at this level and proceed one step further by writing another equation of motion 
for correlation function $\langle{\langle{f_\sigma}{n_{\bar\sigma}}|f_\sigma ^\dag}\rangle\rangle$.
This gives,
\begin{equation}
	\omega \langle{\langle{f_\sigma }{n_{\bar \sigma }|f_\sigma ^\dag }\rangle}\rangle  = \langle \lbrace{f_\sigma } 	{n_{\bar \sigma }},f_\sigma ^\dag\rbrace\rangle + \langle{\langle[{f_\sigma }{n_{\bar \sigma },H]|f_\sigma ^\dag }\rangle}\rangle.
\end{equation}
Let us first focus on the simplest case which is the top-site case. In this case,
calculation of the (anti) commutation relations needed in the above equation we find:
\begin{multline}
	({\omega  - {\varepsilon _0} - U})\langle {\langle {{f_\sigma }{n_{\bar \sigma }}|f_\sigma ^\dag }\rangle } \rangle  = \langle {{n_{\bar \sigma }}}\rangle \\ + V\sum\limits_\vk\langle\langle {{b_{\vk\sigma }}{n_{\bar \sigma }} + {f_\sigma }({f_{\bar \sigma }^\dag {b_{p\bar \sigma }} - b_{p\bar \sigma }^\dag {f_{\bar \sigma }}})|f_\sigma ^\dag } \rangle\rangle.
 \label{eqom1}
\end{multline}
At this stage the decoupling scheme can be applied to the correlation function 
on the right hand side of Eq.~\eqref{eqom1} as follows:
\begin{multline}
\langle\langle {{b_{\vk\sigma }}{n_{\bar \sigma }} + {f_\sigma }({f_{\bar \sigma }^\dag {b_{p\bar \sigma }} - b_{p\bar \sigma }^\dag {f_{\bar \sigma }}})|f_\sigma ^\dag } \rangle\rangle \approx  \\ \langle{n_{\bar \sigma }}\rangle \langle\langle {b_{\vk\sigma }}|{f_\sigma ^\dag }\rangle\rangle
+ \langle{f_{\bar \sigma }^\dag {b_{p\bar \sigma }} - b_{p\bar \sigma }^\dag {f_{\bar \sigma }}}\rangle\langle\langle {f_\sigma }|{f_\sigma ^\dag } \rangle\rangle
\end{multline}
The above decoupling has been performed in such a way to avoid off-diagonal correlation
functions in the spin-indices. Hence the operators carrying $\bar{\sigma}$ spin index
are taken out of the correlations in the form of expectation values~\cite{Hubbard1963}.
Finally to close the set of equations of motion,  we write the equation of motion for 
$\langle\langle {b_{\vk\sigma }}|{f_\sigma ^\dag }\rangle\rangle $, which gives,
\begin{equation}
	\langle {\langle {{b_{p\sigma }}|f_\sigma ^\dag } \rangle } \rangle  = V\frac{\omega}{\omega^2-t^2|\phi(\vk)|^2}\langle {\langle {{f_\sigma }|f_\sigma ^\dag } \rangle } \rangle.
\end{equation}
Therefore at the present approximation, the local Green function becomes,
\begin{equation}
\langle {\langle {{f_\sigma }|f_\sigma ^\dag } \rangle } \rangle  = {\left[ {\omega  - {\varepsilon _0} - 
\Sigma^{\Delta}_d ( \omega  ) - \Sigma '( \omega  )} \right]^{ - 1}},
\label{g2.eqn}
\end{equation}
where $\Sigma^\Delta_d$ describe the hybridization of local electrons with the
band continuum and for the top site situation is given by~\cite{Uchoa}
\be
   \Sigma^\Delta_d(\omega)=-V^2\frac{\omega}{D^2}
   \ln\left(\frac{|\omega^2-D^2|}{\omega^2} \right) -iV^2\frac{\pi|\omega|}{D^2}\theta(D-|\omega|).
   \label{sfs.eqn}
\ee
In the p-wave case as is shown in the appendix, 
within the present approximation, the form of the local
Green function remains the same as Eq.~\eqref{g2.eqn}.
The only difference is that now the hybridization function $\Sigma^\Delta_d$
for hollow site case is given by Eq.~\eqref{selfen}.
Since the Hubbard $U$ acts only locally in the impurity orbital, at this level of approximation,
the self-energy correction coming from the hybridization, $\Sigma^\Delta_d$, and those
coming from Hubbard term, $\Sigma'$, are separable and hence the later turns out to be 
insensitive to the nature of hybridization of the impurity orbital with neighboring orbitals. 
The interaction effects arising from the Hubbard $U$ are given by the
self-energy $\Sigma^\prime(\omega)$ as follows:
\begin{equation}
\Sigma '(\omega) \equiv \frac{{U({\omega  - {\varepsilon _0}})\langle {{n_{\bar \sigma }}} \rangle }}{{\omega  - {\varepsilon _0} - U( {1 - \langle {{n_{\bar \sigma }}} \rangle })}}.
\label{sf2.eqn}
\end{equation}

Note that in this case, unlike the
Hartree approximation where the mere effect of Hubbard $U$ is to shift
$\eps_0$ by $U\langle n_{\bar\sigma}\rangle$, in the present approximation,
the self-energy not only is not of a simple shift form, but also has 
acquired a non-trivial $\omega$-dependence. 

\begin{figure}[tb]
\centering
\includegraphics[width = 8.5cm]{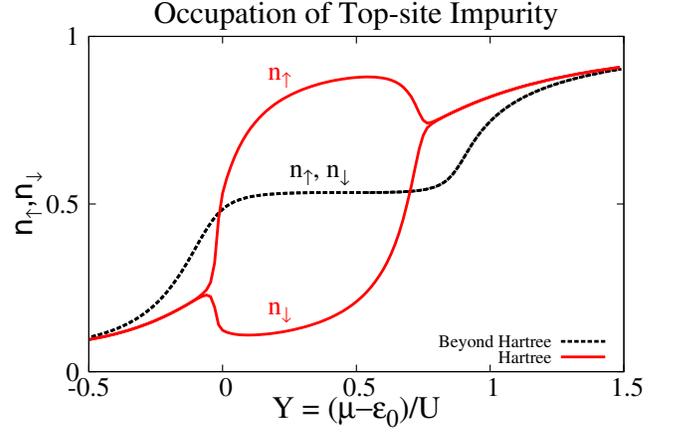}
\caption{ (Color online) 
The self-consistent occupations $n_\up$ and $n_\down$ for top site impurity 
as a function of $Y$ for a fixed value $X=5.0$ and $\eps_0=0.029$. 
As indicated in the legend, within the Hartree approximation
(solid red line), the $n_\up$ and $n_\down$ plots are split, which indicates
the formation of local moments at Hartree level.
But as soon as we go beyond the Hartree approximation (dashed black line),
for the same values of parameters, the $n_\up$ and $n_\down$ curves collapse 
on each other. This means that fluctuations beyond the Hartree approximation
destroy the local moments formed at the mean field level.
}
\label{nbh.fig}
\end{figure}

\begin{figure}[th]
\centering
\includegraphics[width = 8.7cm]{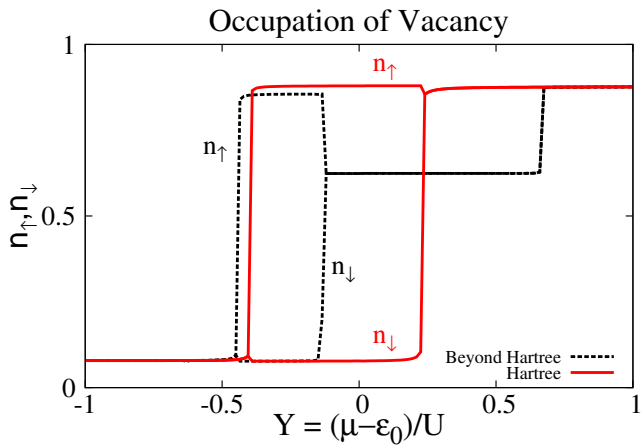}
\caption{ (Color online) The self-consistent occupations $n_\up$ and $n_\down$ 
for hollow site impurity as a function of $Y$ for a fixed value $X=5.0$ and $\eps_0=0.029$.
The region of local moment formation is where the $n_\up$ and $n_\down$ curves split apart. 
The solid line (red) corresponds to Hartree approximation, and the dashed lines (black) 
correspond to post-Hartree approximation. In the p-wave hybridization case 
as can be seen by including the self-energy Eq.~\eqref{sf2.eqn},
the local moment region is shrinked, but it does not vanish. In this sense, local moments
formed by hollow-site impurities are robust against quantum fluctuations.
}
\label{nbh2.fig}
\end{figure}

In Fig.~\ref{nbh.fig} we have compared the self-consistent solutions for the
occupation of the $\up$ and $\down$ spin sub-bands within the Hartree and
our post-Hartree approximation for the top-site impurity. 
The plot has been generated for a fixed 
value of $X=5.0$ and $\eps_0=0.029$ as function of $Y$. Within the Hartree 
approximation plots corresponding to $n_\up$ and $n_\down$ are split for
positive values of $Y$. This indicates the formation of local moments
at the Hartree level. However as soon as we employ the self-energy~\eqref{sf2.eqn}
to go beyond the Hartree, the $n_\up$ and $n_\down$ plots collapse on each other.
Therefore the local moments formed at the Hartree level are destroyed 
by quantum fluctuations beyond the mean field level.

In Fig.~\ref{nbh2.fig} we plot the same set of data as in Fig.~\ref{nbh.fig}
but for the p-wave hybridization function. The interesting feature to note in
this case is that, even by including the self-energy effect 
Eq.~\eqref{sf2.eqn} which goes an step beyond the Hartree approximation, 
still there will be a region in the parameter
space where the local moment survives the fluctuations. 
Therefore the local moments formed by hollow-site impurities (or vacancies) when
compared to those due to top-site impurities, not only are more robust
in terms of the extension of region in the parameter space where 
local moment is formed, but they are also robust against the quantum fluctuation
effects captured by Eq.~\eqref{sf2.eqn}. This makes vacancies and 
hollow-site impurity states ideal sources of local magnetic moments
in graphene which are hard to destroy. This can account for the 
observed magnetic moments in vacant graphene~\cite{Nair}.

\section{Summary and discussion}
In this paper we investigated the formation of local moments in a graphene sheet,
in two situations corresponding to s-wave and p-wave functional dependence in
the hybridization function $V(\vk)$ within the single impurity Anderson model.
The former corresponds to the top-site impurities, while the later may
correspond to hollow-site impurities or vacancies.
First we noted that the output of the self-consistency cycle is very sensitive
to the precision of the integrations. Any approximation in the integrations
propagates the errors thorough the self-consistency cycle and may give rise to
different results. 
Then we compared the Hartree mean field level phase diagram of the
above two cases. We found that vacancies and hollow-site impurities give rise to 
much larger region in the phase space where the magnetic moments are formed. 
In this sense the local moments
arising from impurity's orbitals on the hollow-sites or from vacancies
are robust against change in the parameters. 
Further we showed that within the Hartree mean field the internal 
structure of the impurity orbital itself does not affect the local moment properties.
Then within the equation of motion method, we proceeded one step 
beyond Hartree approximation
and derived a self-energy due to Hubbard $U$ term which takes into 
account quantum fluctuations beyond the Hartree mean field. 
Within such a post-Hartree mean field we obtained the self-consistently
determined occupation numbers of the $\up$ and $\down$ spin impurity
sub-bands. We found that in the top-site case, the quantum fluctuations
in the post-Hartree approximation can destroy the local moments formed
at Hartree level, while in the case of vacancies and hollow-site impurities, 
for the same set of parameters, the local moments 
survive the quantum fluctuations within our approximation.
Therefore we conclude that the p-wave hybridization functions
gives rise to local moments which are not only robust in parameter space,
but are also immune to quantum fluctuations beyond the Hartree mean field.
The present stability analysis is consistent with the observed spin-half 
magnetic states in vacant graphene~\cite{Nair}.

\section{Acknowledgement}
We thank T. Tohyama for useful discussions.
This research was completed while the SAJ was
visiting Yukawa Institute for Theoretical Physics
by the fellowship S13135 from Japan Society for Promotion of Science.

%
%

\begin{appendix}
\section{Beyond Hartree: hollow-site case}
\label{appendixA}
In this appendix we calculate the impurity Green function one step beyond the 
Hartree mean-field approximation for hollow-site impurity. 
The equation of motion for local Green function is given by Eq.~\eqref{eqom}. 
Avoiding the decoupling of 
$\langle\langle n_{\vk\bar{\sigma}}f_{\sigma}|f^\dagger_{\sigma}\rangle\rangle$
and writing an equation of motion for it we obtain,
\begin{equation}
	\omega \langle{\langle{f_\sigma }{n_{\bar \sigma }|f_\sigma ^\dag }\rangle}\rangle  = \langle {n_{\bar \sigma }}\rangle + \langle{\langle[{f_\sigma }{n_{\bar \sigma },H]|f_\sigma ^\dag }\rangle}\rangle,
\end{equation}
which after evaluation of the necessary commutation relations becomes,
\begin{multline}
	({\omega  - {\varepsilon _0} - U})\langle {\langle {{f_\sigma }{n_{\bar \sigma }}|f_\sigma ^\dag }\rangle } \rangle  = \langle {{n_{\bar \sigma }}}\rangle \\ + V\sum\limits_\vk\langle\langle {{c_{\vk\sigma }}{n_{\bar \sigma }} + {f_\sigma }({f_{\bar \sigma }^\dag {c_{p\bar \sigma }} - c_{p\bar \sigma }^\dag {f_{\bar \sigma }}})|f_\sigma ^\dag } \rangle\rangle.
 \label{eqom2}
\end{multline}
In above equation the operator $c^{\dagger}_{\vk\sigma} $ is defined by
\be
   c^{\dagger}_{\vk\sigma} \equiv \phi(\vk)b^{\dagger}_{\vk\sigma} + \phi^*(\vk)a^{\dagger}_{\vk\sigma}.
\ee
In the hollow site configuration the impurity hybridizes with canbon atoms from both sublatices. 
Therefore the operator $c_{\vk\sigma}^\dagger$ in Eq.~\eqref{eqom2} plays the same role as 
$b_{\vk\sigma}^\dagger$ in Eq.~\eqref{eqom1}. We apply the same decoupling scheme which gives,
\begin{multline}
\langle\langle {{c_{\vk\sigma }}{n_{\bar \sigma }} + {f_\sigma }({f_{\bar \sigma }^\dag {c_{\vk\bar \sigma }} - c_{\vk\bar \sigma }^\dag {f_{\bar \sigma }}})|f_\sigma ^\dag } \rangle\rangle \approx  \\ \langle{n_{\bar \sigma }}\rangle \langle\langle {c_{\vk\sigma }}|{f_\sigma ^\dag }\rangle\rangle
+ \langle{f_{\bar \sigma }^\dag {c_{\vk\bar \sigma }} - c_{\vk\bar \sigma }^\dag {f_{\bar \sigma }}}\rangle\langle\langle {f_\sigma }|{f_\sigma ^\dag } \rangle\rangle.
\end{multline}
Hence the Green function at this approximatio for hollow-site configuration will be,
\begin{equation}
\left\langle {\left\langle {{f_\sigma }|f_\sigma ^\dag } \right\rangle } \right\rangle  = 
{\left[ {\omega  - {\varepsilon _0} - \Sigma^{\Delta}_d \left( \omega  \right) - \Sigma '\left( \omega  \right)} \right]^{ - 1}}.\\
\end{equation}
In the above equation the correlation effect has been encoded in self energy$\Sigma'$ which 
in both p-wave and s-wave cases is given by Eq.~\eqref{sf2.eqn}. The only difference
between the two possible hybridization functions appears in $\Sigma^\Delta_d$ which
for the s-wave case is given by Eq.~\eqref{sfs.eqn}, while in the p-wave case it 
is given by Eq.~\eqref{selfen}.
\end{appendix}
\end{document}